\documentclass[a4paper]{article}
\usepackage{url}
\usepackage{graphicx}
\usepackage{hyperref}
\usepackage{natbib}
\usepackage{authblk}
\usepackage{orcidlink}
\newcommand\apj{{\it Astrophys. J.}}

\newcommand\apjs{{\it Astrophys. J.Supp.}}

\newcommand\mnras{{\it Mon. Not. R. Astron. Soc.}}

\newcommand\pasj{{\it Publ. of the Astron. Society of Japan}}

\newcommand\pasa{{\it Publications of the Astronomical Society of Australia}}
\title{MPI-Rockstar: a Hybrid MPI and OpenMP Parallel Implementation of
the Rockstar Halo finder}
\author[1]{Tomoyuki Tokuue}
\author[2]{Tomoaki Ishiyama\orcidlink{0000-0002-5316-9171}~\footnote{ishiyama@chiba-u.jp} }
\author[3,4,5]{Ken Osato\orcidlink{0000-0002-7934-2569}}
\author[6]{Satoshi Tanaka\orcidlink{0000-0003-2442-8784}}
\author[7]{Peter Behroozi\orcidlink{0000-0002-2517-6446}}

\affil[1]{\small Department of Applied and Cognitive Informatics, Division of Mathematics and Informatics, Graduate School of Science and Engineering, Chiba University, Chiba 263-8522, Japan}
\affil[2]{Digital Transformation Enhancement Council, Chiba University, Chiba 263-8522, Japan}
\affil[3]{Center for Frontier Science, Chiba University, Chiba 263-8522, Japan}
\affil[4]{Department of Physics, Graduate School of Science, Chiba University, Chiba 263-8522, Japan}
\affil[5]{Kavli Institute for the Physics and Mathematics of the Universe (WPI), The University of Tokyo, Kashiwa, Chiba 277-8583, Japan}
\affil[6]{Centre for Gravitational Physics and Quantum Information, Yukawa Institute for Theoretical Physics, Kyoto University, Kyoto 606-8502, Japan}
\affil[7]{Department of Astronomy, University of Arizona, AZ 85721, USA}

\date{}

\begin{document}
\maketitle

\section*{Summary}\label{summary}

According to the concordance cosmological theory, structure formation
and evolution proceeds hierarchically in the Universe. Smaller-scale
dark matter structures gravitationally collapse first everywhere in the
Universe, then merge into larger-scale structures. Such dense
gravitationally bound structures of dark matter are called halos. Halos
can host smaller halos, so-called subhalos (or substructures).
Cosmological $N$-body simulations are vital in understanding the
formation and evolution of halos and subhalos. Halo/subhalo finders are
a post-processing step to identify those dense structures in the
particle dataset of cosmological $N$-body simulations.

\href{https://github.com/Tomoaki-Ishiyama/mpi-rockstar/}{MPI-Rockstar}~\footnote{\href{https://github.com/Tomoaki-Ishiyama/mpi-rockstar/}{https://github.com/Tomoaki-Ishiyama/mpi-rockstar/}}
is a massively parallel halo finder based on the
\href{https://bitbucket.org/gfcstanford/rockstar/}{Rockstar}~\footnote{\href{https://bitbucket.org/gfcstanford/rockstar/}{https://bitbucket.org/gfcstanford/rockstar/}}
phase-space temporal halo finder code \citep{Behroozi2013}, which is
one of the most extensively used halo finding codes.  Compared to the
original code, parallelized by a primitive socket communication
library, we parallelized it in a hybrid way using the Message Passing
Interface (MPI) and OpenMP, which is suitable for analysis on the
hybrid shared- and distributed-memory environments of modern
supercomputers. This implementation can easily handle the analysis of
more than a trillion particles on more than 100,000 parallel
processes, enabling the production of a huge dataset for the next
generation of cosmological surveys.

Owing to the advance of supercomputing power and highly scalable
parallel gravitational $N$-body codes
\citep[e.g.,][]{Ishiyama2009b, Ishiyama2012, Potter2017, Wang2018, Garrison2021}, 
the number of
particles in recent massive cosmological simulations exceeds a trillion
\citep[e.g.,][]{Potter2017, Ishiyama2021, Wang2022}, posing significant
challenges for halo finding. Several halo finding algorithms have been
suggested \citep{Knebe2013}, and some of
their implementations are publicly available
\citep[e.g.,][]{Knollmann2009, Behroozi2013, Elahi2019, Springel2021}. 
The computational
performance of these implementations differs substantially, and they
have not been uniformly tested yet on the large-scale hybrid shared- and
distributed-memory environments of modern supercomputers. The original
Rockstar is designed to run on distributed-memory environments, however,
data communications between multiple processes are performed by
one-to-one communications using sockets. The main scaling bottleneck in
Rockstar is that, even when the number of sockets is unlimited, a
single-threaded server process has to coordinate between all the worker
processes. This starts to become a bottleneck around 10,000 processors.
Besides, many sockets (file descriptors) are issued simultaneously in
the case of analysis with many processes, complicating analysis on
modern supercomputers because the number of file descriptors issued
simultaneously is normally limited. MPI-Rockstar addresses these issues
and is designed to run on more than 100,000 parallel processes in a
hybrid way using MPI and OpenMP. As new functions to the original
Rockstar code, MPI-Rockstar supports HDF5 as an output format and can
output additional halo properties such as the inertia tensor.
MPI-Rockstar is not intended to replace the original implementation.

\section{Parallelization}\label{parallelization}

As a process parallelization, the original Rockstar divides a simulation
box by the number of parallel processes and assigns each sub-box to each
process. Then, each process performs 3D Friends-of-Friends (FoF) to find
overdense regions, and FoF halos across processes are linked by
communicating boundary regions. Rockstar then performs the subhalo
finding for each FoF halo using 6D phase space information. Data
communications between multiple processes are performed by one-to-one
communications using sockets.

In MPI-Rockstar, we replaced all socket communications in the original
Rockstar with MPI communications, while maintaining compatibility with
the analysis results. Rather than simply using MPI one-to-one
communication, we changed the order of communication and computation to
utilize collective communications and to run efficiently on large
supercomputers. Furthermore, we parallelized MPI-Rockstar in a hybrid
way, where thread parallelization is implemented within each process
using OpenMP. The subhalo finding is parallelized not only on a process
level but also on a thread level, improving the overall performance of
MPI-Rockstar. This hybrid parallel design also reduces the risk of
per-process out-of-memory compared with a flat-MPI configuration. Figure
1 illustrates this parallelization strategy.

\begin{figure}
\centering
\includegraphics[width=\linewidth]{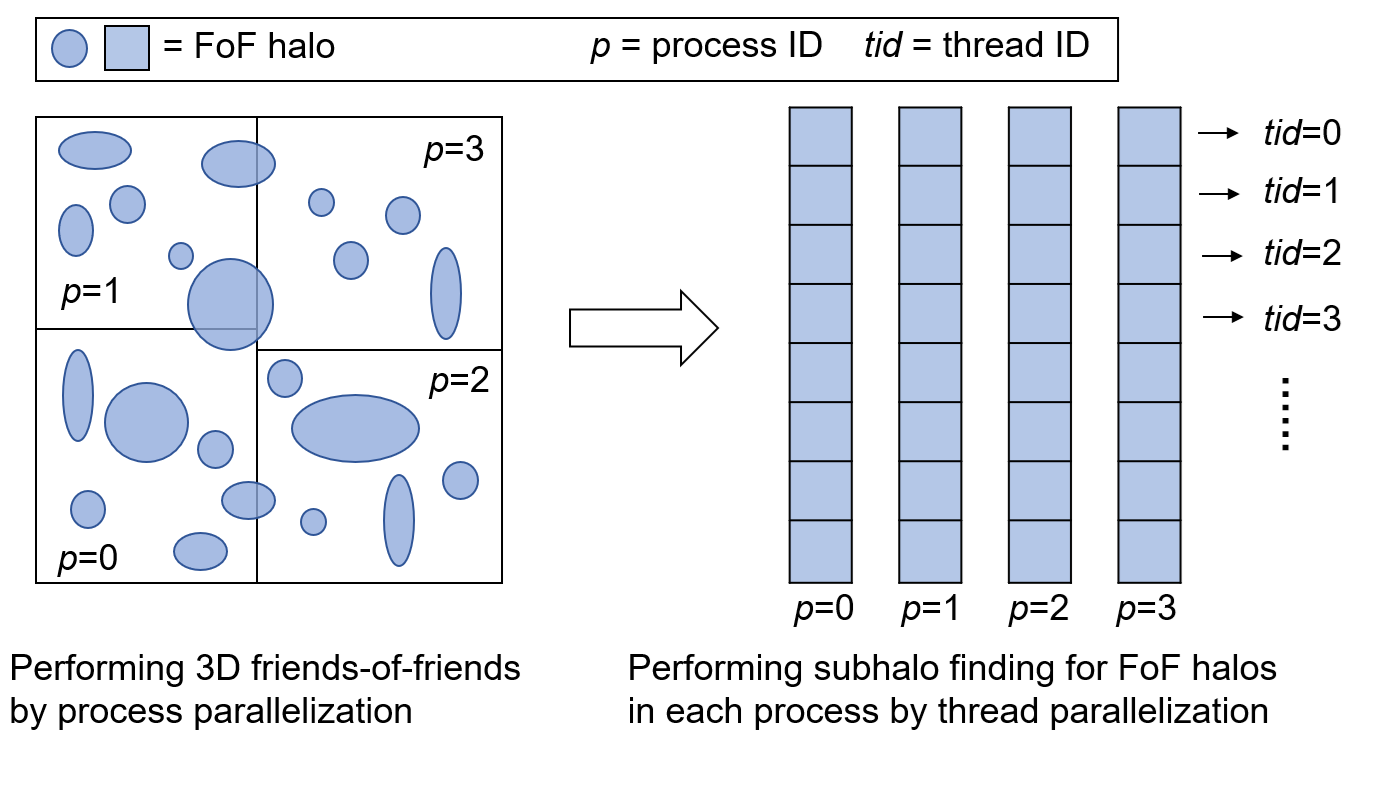}
\caption{Parallelization strategy}
\end{figure}

Figure 2 shows a strong scaling of MPI-Rockstar using up to 1,024 nodes
(48 CPU cores per node) on supercomputer Fugaku. The horizontal and
vertical axes represent the number of computational nodes and the time
taken for the halo and subhalo finding of one snapshot, respectively.
The blue and green curves show the strong scaling for simulations with
$4096^3$ particles in a 2 Gpc/h box and with $2560^3$ particles in a
400 Mpc/h box, respectively. The dotted curves show the ideal scalings.
We analyzed a single snapshot at redshift 2.0. We measured the code's
performance using 2 MPI processes per node and 24 OpenMP threads per
process. This choice gives an optimal configuration for these snapshots,
considering the balance between the computation and I/O time. The
parallel efficiency is excellent, $\sim$90\% for both the
$4096^3$ box from 256 to 1024 nodes and the $2560^3$ box from 256 to
1024 nodes. Thanks to the communication optimization and hybrid
parallelization, MPI-Rockstar could run up to three times faster than
the original Rockstar when compared in the same execution environment.
For example, MPI-Rockstar took $\sim$200 sec for a snapshot at
z=0 with $1024^3$ particles in a 250 Mpc/h box using 64 CPU cores (AMD
EPYC 7452), while the original Rockstar took 440 sec.~Both sets of halo
statistics are consistent with each other. For example, the difference
of halo mass function is below 0.1\% except for the massive and less
massive end, where the statistics are influenced by the small halo
counts and resolution, respectively. Note that Rockstar is not
deterministic, and so identical halo catalogs are generally not possible
to obtain. We also confirm that MPI-Rockstar can analyze 2 trillion
particle simulations on 16,384 nodes (786,432 CPU cores) of Fugaku.

\begin{figure}
%\centering
\includegraphics[width=\linewidth]{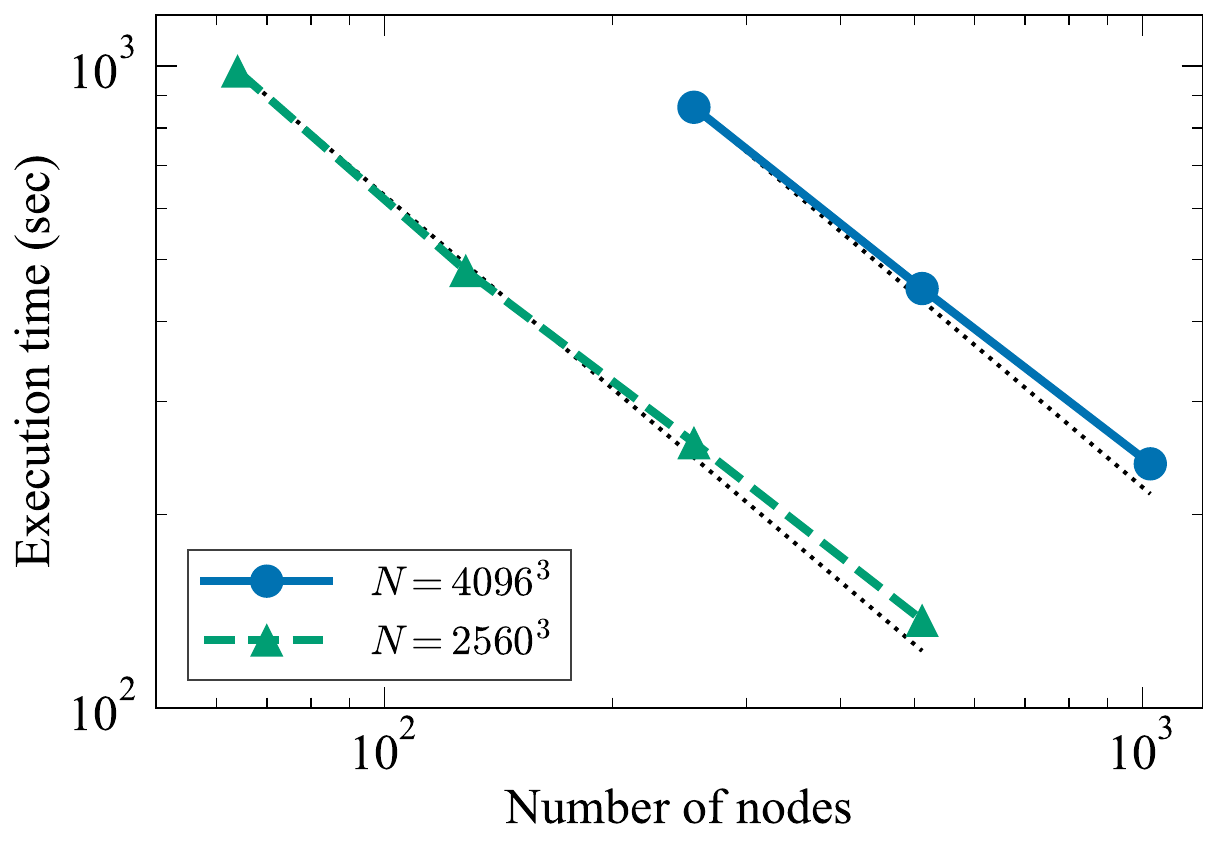}
\caption{Strong scaling of MPI-Rockstar}
\end{figure}

\section{Acknowledgements}\label{acknowledgements}

This work has been supported by IAAR Research Support Program in Chiba
University Japan, MEXT/JSPS KAKENHI (Grant Number JP23H04002), MEXT as
``Program for Promoting Researches on the Supercomputer Fugaku'\,'
(JPMXP1020230406 and JPMXP1020230407), and JICFuS. This work was
partially supported by the Sasakawa Scientific Research Grant from The
Japan Science Society. Numerical computations were carried out on the
supercomputer Fugaku provided by the RIKEN Center for Computational
Science (Project ID: hp230173, and hp240184).

\bibliographystyle{aasjournal}

\end{document}